\documentclass[prl,amsmath,showpacs,draft,twocolumn,superscriptaddress]{revtex4}

\usepackage[final]{graphicx}
\usepackage{bm}
\begin{document}

\title{Dilute Fermi gas in quasi-one-dimensional traps:
  From weakly interacting fermions via hard core bosons to weakly
  interacting Bose gas}

\author{I.~V.~Tokatly}
\email{ilya.tokatly@physik.uni-erlangen.de}

\affiliation{Lerhrstuhl f\"ur Theoretische Festk\"orperphysik,
  Universit\"at Erlangen-N\"urnberg, Staudtstrasse 7/B2, 91058
  Erlangen, Germany}

\affiliation{Moscow Institute of Electronic Technology,
 Zelenograd, 124498 Russia}
\date{\today}

\begin{abstract}

We study equilibrium properties of a cold two-component Fermi gas
confined in a quasi-one-dimensional trap of the transverse
size $l_{\perp}$. In the dilute limit ($nl_{\perp}\ll 1$, where $n$ is
the 1D density) the problem is exactly solvable for an arbitrary 3D
fermionic scattering length $a_{F}$. When $l_{\perp}/a_{F}$ goes from
$-\infty$ to $+\infty$, the system successively passes three
regimes: weakly interacting Fermi gas, hard core Bose gas and
weakly coupled Bose gas. The regimes are separated by two crossovers
at $a_{F}\sim \pm nl_{\perp}^{2}$. In conclusion we discuss experimental
implications of these results.
\end{abstract}


\maketitle 
Trapped cold atomic gases
\cite{Nobel2002} offer a unique possibility to
create experimentally various many-body systems that for a long
time have been considered as purely theoretical models. In this
respect the study of
quasi-one-dimensional (quasi-1D) systems is especially promising
since 1D many-body 
problems are frequently exactly solvable \cite{Mattis}. Currently
quasi-1D cold atomic Bose gases are indeed extensively studied
both experimentally \cite{1dBose-exp} and theoretically
\cite{Olshanii1998,BerMooOls2003,Petr2000+Astrakh2004}.
The behavior of two-component Fermi systems is expected be more diverse
due to one more "degree of freedom" related to the
formation of two-particle composite bosons (dimers). In quasi-3D
traps both Bose-Einstein condensation (BEC) of dimers
\cite{FermiBEC} and the regime of a
crossover from Bardeen-Cooper-Schriefer (BCS) superfluidity to
molecular BEC \cite{Crossover-exp} have been
recently reproduced experimentally. While theoretical studies of BCS-BEC 
crossover in 3D (or quasi-3D) systems have a long history 
(see
Refs.~\onlinecite{Randeria,GTokJETP1993:e,PisStr1996,StiZwe1997,PieStr2000,PerPieStr2003}
and references therein), a similar problem for a strongly anisotropic
confinement has non been addressed up to now. In this paper we
present a complete theory of ``BCS-BEC'' transformation from weakly coupled
Fermi gas to Bose gas of diatomic molecules in quasi-1D traps.

\begin{figure}
   \includegraphics[width=0.48\textwidth]{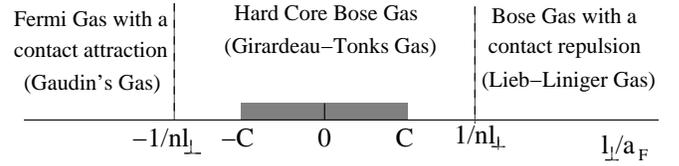}
   \caption{Phase diagram of quasi-1D Fermi gas.}
   \label{fig1}
 \end{figure}

We consider a dilute two-component Fermi gas confined in $(x,y)$
plane by a harmonic potential with characteristic scale
$l_{\perp}=(m\omega_{\perp})^{-1/2}$. In the present context the
diluteness means that $nl_{\perp}\ll 1$, where $n$ is the density
averaged over transverse directions. As usual, assuming that the
interaction potential $V({\bf r})$ is of short range $R_W\ll
l_{\perp}$, we employ the standard pseudopotential approximation:
$V({\bf r})=\frac{4\pi a_{F}}{m}\delta({\bf r})
\frac{\partial}{\partial{r}}r$, where $a_{F}$ is the 3D scattering
length. Parameter $l_{\perp}/a_{F}$ ranges from $-\infty$ to
$+\infty$, which corresponds an attractive interaction and
includes the unitarity limit ($a_{F}=\infty$). We will show that
in the whole range of $a_{F}$ this system is described by exactly
solvable models. At small negative $a_{F}$ ($l_{\perp}/a_{F}\ll
-1$) it behaves according to Gaudin's solution
\cite{Gaudin1967} for an attractive 1D Fermi gas. At $a_{F}\sim
-nl_{\perp}^2$ the Gaudin's Fermi gas transforms to
Girardeau-Tonks gas of hard core bosons
\cite{Girardeau1960,Tonks1936}, which, at positive
$a_{F}>nl_{\perp}^{2}$, smoothly matches the regime of Lieb-Liniger
Bose gas \cite{LieLin1963} with weak repulsion. A schematic phase
diagram is shown in Fig.~\ref{fig1}.
Two crossovers in Fig.~\ref{fig1} reflect two different physical
phenomena. The left crossover corresponds to the transformation from
strongly overlapped BSC-like pairs to well defined composite
bosons, which is roughly similar to the common
BCS-BEC crossover in 3D systems. The right crossover is inherent
for quasi-1D geometry. It is related to a change of boson-boson
interaction due to a  ``dimensional transformation'' from
quasi-1D to 3D character of composite bosons.

Many-body physics of an attractive Fermi gas is closely related to
the solution of a two-body problem. The two-body scattering
problem in an axially symmetric harmonic trap has been already
addressed by several authors
\cite{Olshanii1998,BerMooOls2003,BolTieJul2003}. Below we briefly
discuss properties of a two-body bound state that is a localized
solution to Schr\"odinger equation for the relative motion. The
energy of this state should have a form $E_{0} = \omega_{\perp} -
\varepsilon_{0}$, where $\varepsilon_{0}>0$ is the binding energy.
In the pseudopotential approximation the bound state wave
function, $\chi_{0}({\bf r})$, is expressed in terms of Green's
function, $G(E,{\bf r},{\bf r}')$, of a cylindrical harmonic
oscillator with the mass $m/2$:
\begin{equation}
\chi_{0}({\bf r}) = AG_{0}({\bf r},0),
\label{1}
\end{equation}
Here $G_{0}({\bf r},0)\equiv G(E_{0},{\bf r},0)$, and  $A$ is the
normalization constant which is given by the expression
 \begin{equation}
 A^{2} =\frac{32\pi}{\sqrt{2}m^{2}l_{\perp}}\left[
 \zeta\left(\frac{3}{2},\frac{\varepsilon_{0}}{2\omega_{\perp}}\right)
 \right]^{-1}
 \label{10}
 \end{equation}
where $\zeta(z,\Omega)$ is the zeta function \cite{G&R1965}.
The binding energy $\varepsilon_{0}$ is a solution to the
following transcendental equation
\begin{equation}
\frac{m}{4\pi a_{F}} = \left[\frac{\partial}{\partial{r}}r
G_{0}({\bf r},0)\right]_{r=0}.
\label{2}
\end{equation}
Equation (\ref{2}), which defines a pole in the two-body scattering
amplitude, can be reduced to the form \cite{BerMooOls2003}
\begin{equation}
{l_{\perp}}/{a_{F}}= -
\zeta(1/2,\varepsilon_{0}/2\omega_{\perp})/\sqrt{2}
\label{5}
\end{equation}  
Right hand side in Eq.~(\ref{5}) is a monotonically increasing
function of $\Omega = \varepsilon_{0}/2\omega_{\perp}$.
It diverges as $-1/\sqrt{2\Omega}$ at $\Omega\ll 1$, crosses zero
at $\Omega\approx 0.3$, and  goes to $+\infty$ as $\sqrt{2\Omega}$
at $\Omega\gg 1$. This behavior translates to the monotonic
dependence of the binding energy $\varepsilon_{0}$ on $a_{F}$. For
small negative $a_{F}$ ($l_{\perp}/a_{F}\ll -1$) we get the result
$\varepsilon_{0}\approx a_{F}^{2}/ml_{\perp}^{4}$. In this regime
$\varepsilon_{0}\ll \omega_{\perp}$ which means that the transverse motion
of particles is confined to the lowest state of spatial
quantization. The bound state is strongly anisotropic with the axial
size $\sim 1/\sqrt{m\varepsilon_{0}}\gg l_{\perp}$. In the unitarity
limit,  $l_{\perp}/a_{F}=0$, the  bound state wave function becomes
almost spherically symmetric, while the
energy takes a universal form $\varepsilon_{0}=0.6\omega_{\perp}$.
In the regime of strong attraction ($l_{\perp}/a_{F}\gg 1$)
the binding energy approaches the usual 3D expression: $\varepsilon_{0}\approx
1/ma_{F}^{2}$. 

We also introduce an effective size of the bound state
$a_{0}=1/\sqrt{m\varepsilon_{0}}$, which has the following
behavior in the characteristic regions of $l_{\perp}/a_{F}$
\begin{eqnarray}
a_{0} &\approx& l_{\perp}^{2}/|a_{F}| \gg l_{\perp},
\quad l_{\perp}/a_{F}\ll -1,
\label{7}\\
a_{0} &\approx& 1.29 l_{\perp},
\qquad \qquad l_{\perp}/a_{F} =0,
\label{8}\\
a_{0} &\approx& a_{F} \ll l_{\perp},
\qquad \quad l_{\perp}/a_{F}\gg 1.
\label{9}
\end{eqnarray}

Let us turn to the many-body problem in the limit $nl_{\perp}\ll
1$. We start from the regime of weak attraction: $a_{F}<0$ and
$|a_{F}|\ll l_{\perp}$ (the region to the left of the shaded
region in the phase diagram of Fig.~\ref{fig1}). In this case the
transverse motion of each Fermi particle is confined to the lowest
oscillator's state. Therefore an effective 1D many-body
Hamiltonian is obtained from the original one by the simple
averaging over transverse directions
\begin{equation}
H_{F} = \int dz\Big[\sum_{j=1}^{2}
-\psi_{j}^{\dag}\frac{\partial_{z}^{2}}{2m}\psi_{j}
-g_{F}\psi_{1}^{\dag}\psi_{2}^{\dag}\psi_{2}\psi_{1}\Big],
\label{11}
\end{equation}
where $\psi_{j}(z)$ are the Fermi field operators, and the 1D
fermionic coupling constant takes the form
\begin{equation}
g_{F}=2|a_{F}|/ml_{\perp}^{2}.
\label{12}
\end{equation}
Solving the 1D two-particle problem with the point attractive
interaction of Eq.~(\ref{12}) we find the bound state energy
$\varepsilon_{0}=mg_{F}^{2}/4=a_{F}^{2}/ml_{\perp}^{4}$, which
recovers the corresponding solution to Eq.~(\ref{5}).

The problem of 1D spin-1/2 fermions with an attractive
interaction, Eq.~(\ref{11}), is exactly solvable
\cite{Gaudin1967}. The ground state energy $E_{F}$ is given by the
set of Gaudin's integral equations \cite{Gaudin1967}
\begin{eqnarray}
E_{F} &=& -\frac{1}{8}mg_{F}^{2}n +
2\int_{-K_{0}}^{K_{0}}\frac{k^{2}}{2m}f(k)\frac{dk}{2\pi},
\label{13}\\
f(k) &=& 2 - \int_{-K_{0}}^{K_{0}}
\frac{2mg_{F}}{(mg_{F})^{2} + (k-k')^{2}}f(k')\frac{dk'}{2\pi},
\label{14}\\
n &=& 2\int_{-K_{0}}^{K_{0}}f(k)\frac{dk}{2\pi}.
\label{15}
\end{eqnarray}
The solution of Eqs.~(\ref{13})-(\ref{15}) is governed by
dimensionless parameter $\gamma_{\text{G}}=mg_{F}/n\sim 1/na_{0}
\sim |a_{F}|/nl_{\perp}^{2}$. Since $\gamma_{\text{G}}$ is a ratio
of two small parameters ($|a_{F}|/l_{\perp}$ and $nl_{\perp}$
respectively) it can take any value in the region of applicability
of Eqs.~(\ref{11})-(\ref{15}).

In the weak coupling limit, $\gamma_{\text{G}}\ll 1$
($l_{\perp}/a_{F}\ll -1/nl_{\perp}$), we get a weakly interacting
two-component 1D Fermi gas with the energy
$E_{F}^{0}=\pi^{2}n^{3}/24m$. To reveal the physics behind the
strong coupling limit, $\gamma_{\text{G}}\gg 1$
($l_{\perp}/a_{F}\ll -1/nl_{\perp}$), it is instructive to rescale
all momenta, $k\to p/2$, in Eqs.~(\ref{13})-(\ref{15}) and rewrite
them as follows
\begin{eqnarray}
E_{F} &=& -\varepsilon_{0}\frac{n}{2} +
\int_{-P_{0}}^{P_{0}}\frac{p^{2}}{2M}\widetilde{f}(p)\frac{dp}{2\pi},
\label{16}\\
\widetilde{f}(p) &=& 1 - \int_{-P_{0}}^{P_{0}}
\frac{2Mg_{F}}{(Mg_{F})^{2} + (p-p')^{2}}\widetilde{f}(p')\frac{dp'}{2\pi},
\label{17}\\
\frac{n}{2} &=& \int_{-P_{0}}^{P_{0}}\widetilde{f}(p)\frac{dp}{2\pi},
\label{18}
\end{eqnarray}
where $\widetilde{f}(p)=2f(p/2)$, and $M=2m$. Obviously, the first
term in Eq.~(\ref{16}) is the energy of $n/2$ noninteracting bound
states (composite bosons). The rest of the set of
Eqs.~(\ref{16})-(\ref{18}) coincides (except for the sign of the
integral term in Eq.~(\ref{17})) with a set of Lieb-Liniger
equations \cite{LieLin1963} for a gas of interacting Bose
particles with the mass $M$ and the density $n/2$. This analogy
was already noted in the early paper by Gaudin \cite{Gaudin1967}.
In the strong coupling limit the integral term in Eq.~(\ref{17})
(with a ``wrong'' sign) is irrelevant. Hence in this regime the
system behaves as a gas of impenetrable composite bosons
(Girardeau-Tonks gas). The strong repulsion is a direct
consequence of Fermi statistics and 1D kinematics -- the composite
particles can not go around while the Pauli principle forbids them
to penetrate each other. The energy in this regime is given by the
formula for a one-component Fermi gas of the density $n/2$:
$E_{F}^{\infty}=-\varepsilon_{0}n/2 + \pi^{2}(n/2)^{3}/6M$. A
crossover from the weakly coupled Fermi gas to the Tonks gas is
located at $l_{\perp}/a_{F}\sim -1/nl_{\perp}$. The description of
our system in terms of Gaudin's Hamiltonian, Eq.~(\ref{11}), with
$g_{F}$ of Eq.~(\ref{12}) is valid in the region
$l_{\perp}/a_{F}<-C$, where constant $C$ satisfies the condition
$1/nl_{\perp}\gg C\gg 1$.

Another asymptotic regime corresponds to $a_{F}>0$ and $a_{F}\ll
l_{\perp}$ (the region to the right of the shaded region in
Fig.~\ref{fig1}). This is the regime of small 3D
composite bosons with the size $a_{0}=a_{F}$ (see Eq.~(\ref{9})), the
binding energy $\varepsilon_{0}=1/ma_{F}^{2}$ and 3D scattering
length $a_{B}\approx 0.6a_{F}$
\cite{PetSalShl2003,TokCond-mat2004a}. Since $nl_{\perp}\ll 1$,
the transverse center-of-mass motion of dimers corresponds to the
lowest harmonic oscillator state. Hence the effective 1D bosonic
Hamiltonian reduces to the form
\begin{equation}
H_{B} = -\varepsilon_{0}\frac{n}{2} + \int dz\Big[
-\varphi^{\dag}\frac{\partial_{z}^{2}}{2M}\varphi
+\frac{1}{2}g_{B}\varphi^{\dag}\varphi^{\dag}\varphi\varphi\Big]
\label{19}
\end{equation}
with the following 1D bosonic coupling constant
\begin{equation}
g_{B}=2a_{B}/ml_{\perp}^{2} \approx 1.2 a_{F}/ml_{\perp}^{2}.
\label{20}
\end{equation}
The energy of $n/2$ bosons with a repulsive contact interaction is
given by the exact solution due to Lieb and Liniger
\cite{LieLin1963} (Lieb-Liniger integral equations are equivalent
to Eqs.~(\ref{13})-(\ref{15}) with the replacement $g_{F}\to
-g_{B}$).

A dimensionless parameter that governs the behavior of the system
in this regime is $\gamma_{LL}=Mg_{B}/n\sim a_{F}/nl_{\perp}^{2}$.
In the region $l_{\perp}/a_{F}\gg 1/nl_{\perp}$ (which corresponds
to $\gamma_{LL}\ll 1$) we have a weakly coupled Bose gas with
the energy $E_{B}^{0}=-\varepsilon_{0}n/2 +
\frac{1}{2}g_{B}(n/2)^{2}$ \cite{LieLin1963}. In the opposite
limit $l_{\perp}/a_{F}\ll 1/nl_{\perp}$ ($\gamma_{LL}\gg 1$) we
again recover the hard core 1D bosons, and thus get the energy
$E_{B}^{\infty}=-\varepsilon_{0}n/2 + \pi^{2}(n/2)^{3}/6M$. A
crossover between these two regimes corresponds to
$l_{\perp}/a_{F}\sim 1/nl_{\perp}$.

\begin{figure}
   \includegraphics[width=0.2\textwidth]{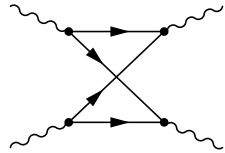}
   \caption{Diagram for the Born scattering amplitude of two composite
   bosons.}
   \label{fig2}
 \end{figure}

Inside the shaded region in Fig.~\ref{fig1} both above asymptotic
approaches (Eqs.~(\ref{11})-(\ref{12}) and
Eqs.~(\ref{19})-(\ref{20})) are not applicable. In this region the
relative motion two particles is neither 1D (as assumed in
Eqs.~(\ref{11})-(\ref{12})) nor 3D (which is the condition for
applicability of Eq.~(\ref{20})). We have, however, seen that on
either side of the shaded region the system behaves a gas of hard
core composite bosons. Below we prove that this model is also
valid everywhere inside. The main point is that for all
$l_{\perp}/a_{F}\gg -1/nl_{\perp}$, which  also covers the
questionable region, the bound state's size $a_{0}$,
Eqs.~(\ref{7})-(\ref{9}), is much smaller than mean interparticle
distance $1/n$. This allows 
us to apply the functional integral approach that has been originally
developed to describe the molecular limit of BCS-BEC crossover in homogeneous
systems 
\cite{Randeria,GTokJETP1993:e,PisStr1996}. The
generalization of this theory to arbitrary quasi-3D trapped systems is given in
Ref.~\onlinecite{TokCond-mat2004a}. Since in harmonic traps the
center-of-mass and the relative motions are decoupled, the general
formalism of Ref.~\onlinecite{TokCond-mat2004a} is directly
applicable to the present quasi-1D problem. The result of this
approach is quite simple and physically appealing -- to the
leading order in $a_{0}n\ll 1$ the system is still described by
the bosonic Hamiltonian of Eq.~(\ref{19}), but $g_{B}$ in general
has a more complicated form. In the present context it is enough
to consider a Born approximation to the coupling constant $g_{B}$,
which corresponds to an ``exchange'' process shown in
Fig.~\ref{fig2} \cite{PieStr2000,TokCond-mat2004a}.
An analytic expression for this diagram takes a form
\begin{equation}
g_{B} = \Lambda_{0}^{4}\int
t_{B}({\bm \rho}_{1},{\bm \rho}_{2},{\bm \rho}_{3},{\bm \rho}_{4})
\prod_{i=1}^{4}\Phi_{0,0}({\bm \rho}_{i})d^{2}{\bm \rho}_{i},
\label{21}
\end{equation}
where ${\bm \rho}=(x,y)$ is the transverse coordinate, $\Phi_{0,0}({\bm \rho})=
l_{\perp}^{-1}\sqrt{2/\pi}e^{-(\rho/l_{\perp})^{2}}$ is the wave 
function for the transverse motion of the center-of-mass, and
$\Lambda_{0}=\int V({\bf r})\chi_{0}({\bf r})d^{3}{\bf r}\equiv A$
(see Eqs.~(\ref{1}), (\ref{10})) is the boson-fermion ``vertex''
\cite{TokCond-mat2004a}. The four-point function $t_{B}$ is
defined as follows
\begin{eqnarray}\nonumber
t_{B}&=&\int\frac{d\omega d k}{(2\pi)^{2}}
{\cal G}_{\omega,k}({\bm \rho}_{1},{\bm \rho}_{2})
{\cal G}_{\omega,k}^{*}({\bm \rho}_{2},{\bm \rho}_{3})\\
&&\times {\cal G}_{\omega,k}({\bm \rho}_{3},{\bm \rho}_{4})
         {\cal G}_{\omega,k}^{*}({\bm \rho}_{4},{\bm \rho}_{1}),
\label{22}
\end{eqnarray}
Here ${\cal G}_{\omega,k}({\bm \rho},{\bm \rho}')$ is the one
particle Green's
function for a harmonically confined Fermi gas with the chemical
potential $\mu = \omega_{\perp}-\varepsilon_{0}/2$
\begin{equation}
{\cal G}_{\omega,k}=\sum_{m=-\infty}^{\infty}\sum_{n=0}^{\infty}
\frac{\phi_{n,m}({\bm \rho})\phi_{n,m}^{*}({\bm \rho})}{i\omega
- \frac{k^{2}}{2m} - \omega_{\perp}(2n+|m|)-\frac{\varepsilon_{0}}{2}}
\label{23}
\end{equation}
where $\phi_{n,m}({\bm \rho})$ are the eigen
functions of a 2D harmonic oscillator with frequency $\omega_{\perp}$.
Reducing all integrals in Eqs.~(\ref{21}), (\ref{22}) to a
dimensionless form we find that $g_{B}$ has the following general
structure
\begin{equation}
g_{B} =
\widetilde{g}(\varepsilon_{0}/\omega_{\perp})/ml_{\perp}
\label{24}
\end{equation}
Dimensionless function $\widetilde{g}(x)$ increases as $6\sqrt{x}$
at $x\ll 1$, reaches a maximum at $x\sim 1$, and goes to zero as
$4/\sqrt{x}$ at $x\gg 1$. Using the known dependence of
$\varepsilon_{0}$ on $l_{\perp}/a_{F}$ we get the required function
$g_{B}(a_{F})$ in all characteristic regimes: $g_{B}\sim
|a_{F}|/ml_{\perp}^{2}$ if $l_{\perp}/|a_{F}|\gg 1$, and
$g_{B}\sim 1/ml_{\perp}$, if $l_{\perp}/|a_{F}|<1$ \cite{note1}.
Renormalization of the Born scattering amplitude does not alter
this general behavior since it can change only numerical
coefficients in the above formulas. Thus, everywhere in the region
$l_{\perp}/|a_{F}|\ll 1/nl_{\perp}$ the strong coupling condition
$mg_{B}/n\gg 1$ is fulfilled. This proves the correctness of the
complete phase diagram in Fig.~\ref{fig1}.

The above formal results lead to the following physical picture of the
quasi-1D ``BCS-BEC'' transformation that consists of two well defined steps.
The first step is the ``Fermi-Bose'' crossover at 
$l_{\perp}/a_{F}\sim -1/nl_{\perp}$. On the left of this point
we have a ``BCS''-like state with strongly overlaped pairs, while
everywhere on the right the system is composed of well
defined composite bosons. The further physical changes in the system
are related to the evolution of an effective boson-boson
interaction. In the region $-1/nl_{\perp}<l_{\perp}/a_{F}<C$ (from the
left crossover to the right end of the shaded region in
Fig.~\ref{fig1}) the bosons are of quasi-1D form. Their transverse
size equals to the transverse size $l_{\perp}$ of the confining
potential. Therefore two colliding bosons can not avoid each
other while their mutual penetration is forbidden by Pauli
principle. As a result we have Girardeau-Tonks regime. In the region 
$C>l_{\perp}/a_{F}>1/nl_{\perp}$ the composite bosons take the common
3D form with radius $a_{0}\approx a_{F}$. However the transverse
confinement of their center-of-mass motion is still strong -- the
system remains in the hard core gas regime. It should be outlined that
the microscopic reason for the Tonks gas regime is quite different on
opposite sides of the shaded region. Finally when the size of
bosons becomes small enough ($a_{0}\approx a_{F}<nl_{\perp}^{2}$) the
system enters the regime of weakly interacting Bose gas. It is worse
mentioning that in contrast to 3D case nothing special happens in the
unitarity limit  $a_{F}=\infty$ that is simply the middle of the
Giradeau-Tonks regime. In fact this regime is a quasi-1D realization
of a universal behavior when all thermodynamic characteristics are
independent of the fermionic scattering length. An important feature of
quasi-1D systems is that the universal behavior extends to a wide
region around the point $a_{F}=\infty$.

Experimentally quasi-1D atomic gases are normally produced in highly
elongated needle-shaped traps
\cite{1dBose-exp} while the
scattering length $a_{F}$ can be tuned using the Feshbach resonance
technique 
\cite{FermiBEC,Crossover-exp}. 
In such systems the double-crossover structure of the phase diagram can be
observed via changes of the density profile $n(z)$. If the axial confinement
(with a scale 
$l_{z}\gg l_{\perp}$) is semiclassical we can calculate $n(z)$
using the local density approximation. In the regime of weakly
interacting Gaudin's gas we get the result
$n_{\text{G}}(z)=(2/\pi l_{z})\sqrt{N-z^2/l_{z}^2}$, which
corresponds to noninteracting fermions (here $N$ is the total number of
particles). The density distribution
in the Girardeau-Tonks regime is given by the similar
expression $n_{\text{GT}}(z)=(2\sqrt{2}/\pi
l_{z})\sqrt{N-2z^2/l_{z}^2}=\sqrt{2}n_{\text{G}}(\sqrt{2}z)$. Hence
when we scan through the first crossover at $l_{\perp}/a_{F}\sim
-1/n(0)l_{\perp}\sim -\frac{l_{z}}{l_{\perp}\sqrt{N}}$, the cloud
shrinks down while preserving the Fermi-like shape. The density
distribution remains unchanged until the next crossover at
$l_{\perp}/a_{F}\sim \frac{l_{z}}{l_{\perp}\sqrt{N}}$ is reached.
This is the universal or unitarity limited regime as it
includes the point $a_{F}=\infty$. In the region
$l_{\perp}/a_{F}>\frac{l_{z}}{l_{\perp}\sqrt{N}}$ the system
gradually transforms to a weakly coupled Bose gas with a parabolic
density profile 
$n_{B}(z)=n_{B}(0)(1-z^{2}/Z_{\text{TF}}^{2})$, where
$n_{B}(0)= a_{B}^{-1}(l_{\perp}Z_{\text{TF}}/l_{z}^{2})^{2}$ and
$Z_{\text{TF}}= (3a_{B}l_{z}^{4}N/4l_{\perp}^{2})^{1/3}$. The
evolution of $n(z)$ can be visualized in terms of the following
two-step dependence of the mean-square size, $\langle
z^{2}\rangle=N^{-1}\int z^{2}n(z)dz$, on $l_{\perp}/a_{F}$ 
\begin{equation}
\langle z^{2}\rangle  =
 \begin{cases}
 \frac{1}{4}l_{z}^{2}N,
& \frac{l_{\perp}}{a_{F}} < -\frac{l_{z}}{l_{\perp}\sqrt{N}}\\
 \frac{1}{8}l_{z}^{2}N,
& \frac{l_{\perp}}{|a_{F}|} < \frac{l_{z}}{l_{\perp}\sqrt{N}}\\
 \frac{1}{5}l_{z}^{2}\left(\frac{3l_{z}a_{B}N}{4l_{\perp}^{2}}\right)^{2/3},
 & \frac{l_{\perp}}{a_{F}} > \frac{l_{z}}{l_{\perp}\sqrt{N}}.
 \end{cases}
\label{25}
\end{equation}

The most characteristic feature of quasi-1D Fermi gases is a long,
well pronounced plateau with $\langle z^{2}\rangle\approx
\frac{1}{8}l_{z}^{2}N$ in the region
$l_{\perp}/|a_{F}|<\frac{l_{z}}{l_{\perp}\sqrt{N}}$ (see
Eq.~(\ref{25})). This should be 
contrasted to the plain single-crossover structure of $\langle
z^{2}\rangle$ in 3D traps, which has been observed in
Ref.~\cite{Crossover-exp}. Therefore the observation of such a plateau
around the Feshbach resonance in a highly elongated trapped Fermi
system unambiguously indicates a realization of the quasi-1D regime.  


\begin{thebibliography}{34}
\expandafter\ifx\csname natexlab\endcsname\relax\def\natexlab#1{#1}\fi
\expandafter\ifx\csname bibnamefont\endcsname\relax
  \def\bibnamefont#1{#1}\fi
\expandafter\ifx\csname bibfnamefont\endcsname\relax
  \def\bibfnamefont#1{#1}\fi
\expandafter\ifx\csname citenamefont\endcsname\relax
  \def\citenamefont#1{#1}\fi
\expandafter\ifx\csname url\endcsname\relax
  \def\url#1{\texttt{#1}}\fi
\expandafter\ifx\csname urlprefix\endcsname\relax\def\urlprefix{URL }\fi
\providecommand{\bibinfo}[2]{#2}
\providecommand{\eprint}[2][]{\url{#2}}

\bibitem[{\citenamefont{Cornell and Wieman}(2002)}]{Nobel2002}
\bibinfo{author}{\bibfnamefont{E.~A.} \bibnamefont{Cornell}} \bibnamefont{and}
  \bibinfo{author}{\bibfnamefont{C.~E.} \bibnamefont{Wieman}},
  \bibinfo{journal}{Rev.\ Mod.\ Phys.} \textbf{\bibinfo{volume}{74}},
  \bibinfo{pages}{875} (\bibinfo{year}{2002}); 
\bibinfo{author}{\bibfnamefont{W.}~\bibnamefont{Ketterle}},
  {\em ibid}. \textbf{\bibinfo{volume}{74}},
  \bibinfo{pages}{1131} (\bibinfo{year}{2002}).

\bibitem[{\citenamefont{Mattis}(1993)}]{Mattis}
\bibinfo{author}{\bibfnamefont{D.~C.} \bibnamefont{Mattis}},
  \emph{\bibinfo{title}{The Many-Body Problem}} (\bibinfo{publisher}{World
  Scientific}, \bibinfo{address}{Singapore}, \bibinfo{year}{1993}).

\bibitem[{\citenamefont{G{\"o}rlitz et~al.}(2001)}]{1dBose-exp}
\bibinfo{author}{\bibfnamefont{A.}~\bibnamefont{G{\"o}rlitz}},
  \bibnamefont{et~al.}, \bibinfo{journal}{Phys.\ Rev.\ Lett.}
  \textbf{\bibinfo{volume}{87}}, \bibinfo{pages}{130402}
  (\bibinfo{year}{2001});
\bibinfo{author}{\bibfnamefont{S.}~\bibnamefont{Dettmer}},
  \bibnamefont{et~al.}, {\em ibid.}
  \textbf{\bibinfo{volume}{87}}, \bibinfo{pages}{160406}
  (\bibinfo{year}{2001});
\bibinfo{author}{\bibfnamefont{D.}~\bibnamefont{Hellweg}},
  \bibnamefont{et~al.}, {\em ibid.}
  \textbf{\bibinfo{volume}{91}}, \bibinfo{pages}{010406}
  (\bibinfo{year}{2003});
\bibinfo{author}{\bibfnamefont{S.}~\bibnamefont{Richard}},
  \bibnamefont{et~al.}, {\em ibid.}
  \textbf{\bibinfo{volume}{91}}, \bibinfo{pages}{010405}
  (\bibinfo{year}{2003});
\bibinfo{author}{\bibfnamefont{H.}~\bibnamefont{Moritz}}, \bibnamefont{et~al.},
  {\em ibid.} \textbf{\bibinfo{volume}{91}},
  \bibinfo{pages}{250402} (\bibinfo{year}{2003}).

\bibitem[{\citenamefont{Olshanii}(1998)}]{Olshanii1998}
\bibinfo{author}{\bibfnamefont{M.}~\bibnamefont{Olshanii}},
  \bibinfo{journal}{Phys.\ Rev.\ Lett.} \textbf{\bibinfo{volume}{81}},
  \bibinfo{pages}{938} (\bibinfo{year}{1998}).

\bibitem[{\citenamefont{Bergeman et~al.}(2003)\citenamefont{Bergeman, Moore,
  and Olshanii}}]{BerMooOls2003}
\bibinfo{author}{\bibfnamefont{T.}~\bibnamefont{Bergeman}},
  \bibinfo{author}{\bibfnamefont{M.~G.} \bibnamefont{Moore}}, \bibnamefont{and}
  \bibinfo{author}{\bibfnamefont{M.}~\bibnamefont{Olshanii}},
  \bibinfo{journal}{Phys.\ Rev.\ Lett.} \textbf{\bibinfo{volume}{91}},
  \bibinfo{pages}{163201} (\bibinfo{year}{2003}).

\bibitem[{\citenamefont{Petrov et~al.}(2000)\citenamefont{Petrov, Shlyapnikov,
  and Walraven}}]{Petr2000+Astrakh2004}
\bibinfo{author}{\bibfnamefont{D.~S.} \bibnamefont{Petrov}},
  \bibinfo{author}{\bibfnamefont{G.~V.} \bibnamefont{Shlyapnikov}},
  \bibnamefont{and} 
  \bibinfo{author}{\bibfnamefont{J.~T.~M.} \bibnamefont{Walraven}},
  \bibinfo{journal}{Phys.\ Rev.\ Lett.} \textbf{\bibinfo{volume}{85}},
  \bibinfo{pages}{3745} (\bibinfo{year}{2000});
\bibinfo{author}{\bibfnamefont{G.~E.} \bibnamefont{Astrakharchik}},
  \bibnamefont{et~al.}, {\em ibid.}
  \textbf{\bibinfo{volume}{92}}, \bibinfo{pages}{030402}
  (\bibinfo{year}{2004}).

\bibitem[{\citenamefont{Jochim et~al.}(2003)}]{FermiBEC}
\bibinfo{author}{\bibfnamefont{S.}~\bibnamefont{Jochim}}, \bibnamefont{et~al.},
  \bibinfo{journal}{Science} \textbf{\bibinfo{volume}{302}},
  \bibinfo{pages}{2101} (\bibinfo{year}{2003});
\bibinfo{author}{\bibfnamefont{M.}~\bibnamefont{Greiner}},
  \bibinfo{author}{\bibfnamefont{C.~A.} \bibnamefont{Regal}}, \bibnamefont{and}
  \bibinfo{author}{\bibfnamefont{D.~S.} \bibnamefont{Jin}},
  \bibinfo{journal}{Nature} \textbf{\bibinfo{volume}{426}},
  \bibinfo{pages}{537} (\bibinfo{year}{2003});
\bibinfo{author}{\bibfnamefont{M.~W.} \bibnamefont{Zwierlein}},
  \bibnamefont{et~al.}, \bibinfo{journal}{Phys.\ Rev.\ Lett.}
  \textbf{\bibinfo{volume}{91}}, \bibinfo{pages}{250401}
  (\bibinfo{year}{2003}).

\bibitem[{\citenamefont{Bartenstein et~al.}(2004)}]{Crossover-exp}
\bibinfo{author}{\bibfnamefont{M.}~\bibnamefont{Bartenstein}},
  \bibnamefont{et~al.}, \bibinfo{journal}{Phys.\ Rev.\ Lett.}
  \textbf{\bibinfo{volume}{92}}, \bibinfo{pages}{120401}
  (\bibinfo{year}{2004});
\bibinfo{author}{\bibfnamefont{C.~A.} \bibnamefont{Regal}},
  \bibinfo{author}{\bibfnamefont{M.}~\bibnamefont{Greiner}}, \bibnamefont{and}
  \bibinfo{author}{\bibfnamefont{D.~S.} \bibnamefont{Jin}},
  {\em ibid.} \textbf{\bibinfo{volume}{92}},
  \bibinfo{pages}{040403} (\bibinfo{year}{2004});
\bibinfo{author}{\bibfnamefont{M.~W.} \bibnamefont{Zwierlein}},
  \bibnamefont{et~al.}, {\em ibid.}
  \textbf{\bibinfo{volume}{92}}, \bibinfo{pages}{120403}
  (\bibinfo{year}{2004}).

\bibitem[{\citenamefont{{Sa de Melo} et~al.}(1993)\citenamefont{{Sa de Melo},
  Randeria, and Engelbrecht}}]{Randeria}
\bibinfo{author}{\bibfnamefont{C.~A.~R.} \bibnamefont{{Sa de Melo}}},
  \bibinfo{author}{\bibfnamefont{M.}~\bibnamefont{Randeria}}, \bibnamefont{and}
  \bibinfo{author}{\bibfnamefont{J.~R.} \bibnamefont{Engelbrecht}},
  \bibinfo{journal}{Phys.\ Rev.\ Lett.} \textbf{\bibinfo{volume}{71}},
  \bibinfo{pages}{3202} (\bibinfo{year}{1993});
\bibinfo{author}{\bibfnamefont{J.~R.} \bibnamefont{Engelbrecht}},
  \bibinfo{author}{\bibfnamefont{M.}~\bibnamefont{Randeria}}, \bibnamefont{and}
  \bibinfo{author}{\bibfnamefont{C.~A.~R.} \bibnamefont{{Sa de Melo}}},
  \bibinfo{journal}{Phys.\ Rev.\ B} \textbf{\bibinfo{volume}{55}},
  \bibinfo{pages}{15153} (\bibinfo{year}{1997}).

\bibitem[{\citenamefont{Gorbatsevich and Tokatly}(1993)}]{GTokJETP1993:e}
\bibinfo{author}{\bibfnamefont{A.~A.} \bibnamefont{Gorbatsevich}}
  \bibnamefont{and} \bibinfo{author}{\bibfnamefont{I.~V.}
  \bibnamefont{Tokatly}}, \bibinfo{journal}{Sov. Phys. JETP}
  \textbf{\bibinfo{volume}{76}}, \bibinfo{pages}{347} (\bibinfo{year}{1993}),
  \bibinfo{note}{[Zh. Exp. Teor. Fiz. {\bf 103}, 702 (1993)]}.

\bibitem[{\citenamefont{Pistolesi and Strinati}(1996)}]{PisStr1996}
\bibinfo{author}{\bibfnamefont{F.}~\bibnamefont{Pistolesi}} \bibnamefont{and}
  \bibinfo{author}{\bibfnamefont{G.~C.} \bibnamefont{Strinati}},
  \bibinfo{journal}{Phys.\ Rev.\ B} \textbf{\bibinfo{volume}{53}},
  \bibinfo{pages}{15168} (\bibinfo{year}{1996}).

\bibitem[{\citenamefont{Stintzing and Zwerger}(1997)}]{StiZwe1997}
\bibinfo{author}{\bibfnamefont{S.}~\bibnamefont{Stintzing}} \bibnamefont{and}
  \bibinfo{author}{\bibfnamefont{W.}~\bibnamefont{Zwerger}},
  \bibinfo{journal}{Phys.\ Rev.\ B} \textbf{\bibinfo{volume}{56}},
  \bibinfo{pages}{9004} (\bibinfo{year}{1997}).

\bibitem[{\citenamefont{Pieri and Strinati}(2000)}]{PieStr2000}
\bibinfo{author}{\bibfnamefont{P.}~\bibnamefont{Pieri}} \bibnamefont{and}
  \bibinfo{author}{\bibfnamefont{G.~C.} \bibnamefont{Strinati}},
  \bibinfo{journal}{Phys.\ Rev.\ B} \textbf{\bibinfo{volume}{61}},
  \bibinfo{pages}{15370} (\bibinfo{year}{2000}).

\bibitem[{\citenamefont{Perali et~al.}(2003)\citenamefont{Perali, Pieri, and
  Strinati}}]{PerPieStr2003}
\bibinfo{author}{\bibfnamefont{A.}~\bibnamefont{Perali}},
  \bibinfo{author}{\bibfnamefont{P.}~\bibnamefont{Pieri}}, \bibnamefont{and}
  \bibinfo{author}{\bibfnamefont{G.~C.} \bibnamefont{Strinati}},
  \bibinfo{journal}{Phys.\ Rev.\ A} \textbf{\bibinfo{volume}{68}},
  \bibinfo{pages}{031601} (\bibinfo{year}{2003}).

\bibitem[{\citenamefont{Gaudin}(1967)}]{Gaudin1967}
\bibinfo{author}{\bibfnamefont{M.}~\bibnamefont{Gaudin}},
  \bibinfo{journal}{Phys. Lett.} \textbf{\bibinfo{volume}{24A}},
  \bibinfo{pages}{55} (\bibinfo{year}{1967}).

\bibitem[{\citenamefont{Girardeau}(1960)}]{Girardeau1960}
\bibinfo{author}{\bibfnamefont{M.}~\bibnamefont{Girardeau}},
  \bibinfo{journal}{J. Mat. Phys. (N. Y.)} \textbf{\bibinfo{volume}{1}},
  \bibinfo{pages}{516} (\bibinfo{year}{1960}).

\bibitem[{\citenamefont{Tonks}(1936)}]{Tonks1936}
\bibinfo{author}{\bibfnamefont{L.}~\bibnamefont{Tonks}},
  \bibinfo{journal}{Phys.\ Rev.} \textbf{\bibinfo{volume}{50}},
  \bibinfo{pages}{955} (\bibinfo{year}{1936}).

\bibitem[{\citenamefont{Lieb and Liniger}(1963)}]{LieLin1963}
\bibinfo{author}{\bibfnamefont{E.~H.} \bibnamefont{Lieb}} \bibnamefont{and}
  \bibinfo{author}{\bibfnamefont{W.}~\bibnamefont{Liniger}},
  \bibinfo{journal}{Phys.\ Rev.} \textbf{\bibinfo{volume}{130}},
  \bibinfo{pages}{1605} (\bibinfo{year}{1963}).

\bibitem[{\citenamefont{Bolda et~al.}(2003)\citenamefont{Bolda, Tiesinga, and
  Julienne}}]{BolTieJul2003}
\bibinfo{author}{\bibfnamefont{E.~L.} \bibnamefont{Bolda}},
  \bibinfo{author}{\bibfnamefont{E.}~\bibnamefont{Tiesinga}}, \bibnamefont{and}
  \bibinfo{author}{\bibfnamefont{P.~S.} \bibnamefont{Julienne}},
  \bibinfo{journal}{Phys.\ Rev.\ A} \textbf{\bibinfo{volume}{68}},
  \bibinfo{pages}{032702} (\bibinfo{year}{2003}).

\bibitem[{\citenamefont{Gradshteyn and Ryzhik}(1965)}]{G&R1965}
\bibinfo{author}{\bibfnamefont{I.~S.} \bibnamefont{Gradshteyn}}
  \bibnamefont{and} \bibinfo{author}{\bibfnamefont{I.~M.}
  \bibnamefont{Ryzhik}}, \emph{\bibinfo{title}{Tables of Integrals, Series and
  Products}} (\bibinfo{publisher}{Academic Press}, \bibinfo{address}{New York},
  \bibinfo{year}{1965}).

\bibitem[{\citenamefont{Petrov et~al.}(2003)\citenamefont{Petrov, Salomon, and
  Shlyapnikov}}]{PetSalShl2003}
\bibinfo{author}{\bibfnamefont{D.}~\bibnamefont{Petrov}},
  \bibinfo{author}{\bibfnamefont{C.}~\bibnamefont{Salomon}}, \bibnamefont{and}
  \bibinfo{author}{\bibfnamefont{G.}~\bibnamefont{Shlyapnikov}},
  \bibinfo{journal}{cond-mat/0309010}  (\bibinfo{year}{2003}).

\bibitem[{\citenamefont{Tokatly}(2004)}]{TokCond-mat2004a}
\bibinfo{author}{\bibfnamefont{I.~V.} \bibnamefont{Tokatly}},
  \bibinfo{journal}{e-print cond-mat/0401603}  (\bibinfo{year}{2004}).

\bibitem[{not()}]{note1}
\bibinfo{note}{The exact asymptotics of $g_{B}$, Eq.~(\ref{21}), are:
  $g_{B}\approx 6/ma_{0}$ for $l_{\perp}/a_{F}\ll-1$, which recovers the purely
  1D result \cite{GTokJETP1993:e}, and $g_{B}\approx 4a_{F}/ml_{\perp}^{2}$ for
  $l_{\perp}/a_{F}\gg 1$, which is in agreement with the 3D unrenormalized
  bosonic scattering length $a_{B}^{0}=2a_{F}$ (see, for example,
  Ref~\onlinecite{TokCond-mat2004a}).}

\end{thebibliography}

\end{document}